\documentclass{optica-article}

\journal{opticajournal} 

\articletype{Research Article}

\usepackage{lineno}
\usepackage{graphicx}
\usepackage{subcaption}
\usepackage{upgreek}
\usepackage{float}

\begin{document}

\title{Compact fiber-based compression of a 1~W, 76~MHz Yb laser to 15~fs for broadband ultrafast applications}

\author{Marco Polastri,\authormark{1,*} Erika Benedetti,\authormark{1} Eva Arianna Aurelia Pogna,\authormark{2} and Nicola Coluccelli\authormark{1}}

\address{\authormark{1}Dipartimento di Fisica - Politecnico di Milano, Piazza Leonardo da Vinci 32 Milano 20133, Italy\\
\authormark{2}Istituto di
Fotonica e Nanotecnologie, Consiglio Nazionale delle Ricerche (CNR-IFN), 20133 Milano, Italy}

\email{\authormark{*}marco.polastri@polimi.it} 


\begin{abstract*}

We demonstrate a compact scheme for generating sub-20-fs pulses from a commercial ytterbium femtosecond laser delivering 80~fs pulses at 76~MHz repetition rate with 1~W average power. Spectral broadening is achieved in a photonic crystal fiber (PCF), followed by dispersion compensation using broadband chirped mirrors. By systematically varying the fiber length and coupled power, we investigate the interplay between nonlinear spectral broadening and higher-order dispersion. While the spectral bandwidth increases monotonically with fiber length, the achievable pulse duration exhibits a clear minimum due to accumulation of uncompensated higher-order phase, primarily third-order dispersion. An optimal fiber length of 80~mm yields nearly transform-limited 15.4~fs pulses. Shorter fibers provide insufficient broadening, whereas longer fibers, despite offering larger bandwidth, compromise the pulse temporal quality. Stable sub-20-fs operation is demonstrated at average powers exceeding 600~mW, and noise measurements indicate that the system performance is limited by the Yb seed laser. These results identify an optimal nonlinear interaction regime in PCF-based compression and establish a practical design rule linking spectral broadening to higher-order phase for compact ultrafast Yb laser sources.
\end{abstract*}

\section{Introduction}

Ultrashort laser pulses are key tools in modern photonics, enabling time-resolved spectroscopy, nonlinear optics, and coherent control \cite{Backus1998,Keller2010,Tu2013}. Access to few-cycle pulses is particularly important for applications requiring high peak intensity and broad bandwidth, such as ultrafast spectroscopy and frequency conversion \cite{Metzger2011,demmler2011generation,Pupeza2015}. Ti:sapphire Kerr-lens mode-locked oscillators have historically enabled sub-10-fs and even sub-two-cycle pulse generation thanks to their large gain bandwidth \cite{Christov1996,Morgner1999}, but their complexity and lack of robustness limit their practical use.

Ytterbium femtosecond lasers have emerged as a compact and efficient alternative, combining power scalability, high efficiency, robust diode pumping and excellent energy and pointing stability \cite{Zheng2025,Lucca2004,Tian2021,Su2024}. These systems are well suited for high-repetition-rate applications and have enabled significant advances in ultrafast sources and frequency comb technologies \cite{Camenzind23,Hsieh2014,Burghoff2014}. The high repetition rate enables fast data acquisition, improved statistics, and low photodamage, which are crucial for high-sensitivity ultrafast spectroscopy and nonlinear microscopy, particularly in biological applications such as multiphoton deep-tissue imaging, and coherent Raman imaging\cite{Coluccelli2018}. Yb lasers, are also attractive for field resolved applications and terahertz (THz) photonics, where compact and stable drivers are required \cite{Vodopyanov2008,Dietz2014,Hsieh2014,Suerra2025}. However, the narrower gain bandwidth typically limits pulse durations of Yb lasers to $\sim$100~fs \cite{Lucca2004,Tian2021,Su2024}, preventing direct access to the few-cycle regime.

To overcome this limitation, nonlinear spectral broadening followed by external compression is widely employed \cite{druon2004pulse,Lavenu2018,Emaury2014,Mak2015,Fritsch2018}. Photonic crystal fibers (PCFs) are particularly attractive for this purpose due to their high nonlinearity, compactness, and dispersion engineering \cite{druon2004pulse,Tu2013,hooper2011coherent}. Alternative approaches based on hollow-core fibers or multipass cells enable compression at higher pulse energies and average powers \cite{Emaury2014,Mak2015,Lavenu2018}, whereas fiber-based schemes remain advantageous for compact and high-repetition-rate systems.

However, pulse shortening is not solely determined by spectral broadening. Increasing the nonlinear interaction enhances bandwidth, but also leads to the accumulation of higher-order phase distortions. In particular, third-order dispersion and related effects can limit compressibility, introduce temporal pedestals, and prevent full compression \cite{druon2004pulse,Christov1996,demmler2011generation}. As a result, the shortest pulse duration is not necessarily obtained from the widest spectrum.

This limitation is closely related to spectral phase control. Techniques such as multiphoton intrapulse interference phase scan (MIIPS) demonstrate that higher-order phase terms critically affect compression of broadband pulses \cite{Harris2007,Comin2014}. While adaptive compensation can correct these effects, it increases system complexity. From a practical standpoint, it is therefore important to identify conditions that allow near-transform-limited (TL) pulses using simple and robust dispersion management.

This issue is particularly relevant for broadband THz generation by optical rectification, where short optical pulses and stable high-repetition-rate sources are required \cite{Vodopyanov2008,Dietz2014,Suerra2025}. Yb based systems are well suited for these applications, but their native pulse duration is typically too long, motivating the development of compact and efficient compression stages.

In this work, we demonstrate a compact PCF-based compressor for a commercial 1~W Yb femtosecond laser and perform a systematic study of the interplay between nonlinear spectral broadening and higher-order dispersion. By varying the fiber length and coupled power, we identify an optimal nonlinear interaction length arising from the trade-off between spectral broadening and uncompensated higher-order phase, primarily third-order dispersion. This leads to near-transform-limited 15~fs pulses from an initially 80~fs source and establishes a practical design rule for optimizing PCF-based compression in compact Yb laser systems.

\section{Experimental setup and fiber length optimization}

\begin{figure}[t]
\centering
\includegraphics[width=0.8\linewidth]{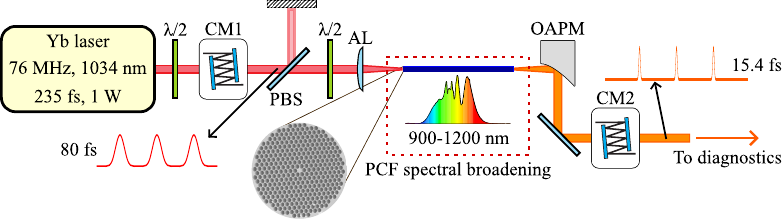}
\caption{Schematic of the compact fiber-based pulse compressor. The output of a 1~W Yb laser source is pre-compressed by a chirped-mirror pair (CM1). Power at the input of the photonic crystal fiber (PCF) is controlled by the combination of a half-wave plate ($\uplambda$/2) and a polarizing beam splitter (PBS). A second half-wave plate is used to align the input electric field polarization with the fiber's principal axis. The beam is focused into the PCF with an aspheric lens (AL) of 15.3~mm focal length for spectral broadening. The spectrally broadened pulses are collimated by an off-axis parabolic mirror (OAPM) and compressed to their nearly transform-limited duration (15.4~fs) by a second pair of chirped mirrors (CM2).}
\label{fig:setup}
\end{figure}

The experimental setup, illustrated in Figure \ref{fig:setup}, is characterized by a simple and compact design. The source consists of a commercial Yb laser (oscillator output of PHAROS, Light Conversion) emitting pulses centered at 1034~nm, with a 235~fs duration, a maximum average power of 1~W, and a repetition rate of 76~MHz. Its 29.6~nm 4-rms bandwidth allows for a TL pulse duration of 80~fs. To achieve the TL duration, a pair of chirped mirrors (Ultrafast Innovations GmbH HD1310) providing a nominal group delay dispersion (GDD) of $-1000~\text{fs}^2$ per bounce, is positioned at the source output. After optimization, the ideal GDD compensation was found to be $-6000~\text{fs}^2$. Following this compression stage, the beam is focused into a PCF with a 10~\textmu m core (Thorlabs LMA-PM-10), which serves as a nonlinear broadening stage to increase the initial bandwidth. To focus the beam (diameter 1/e$^2$ $\approx$1.6~mm) into the PCF, we use an aspheric lens with 15.3~mm focal length and numerical aperture NA= 0.16 (Thorlabs C260TME-C). The power entering the PCF is adjusted using a half-wave plate ($\uplambda$/2, Thorlabs WPH05M) followed by a polarizing beamsplitter (PBS, Thorlabs PBSW1030). A second half-wave plate is placed before the PCF to align the input electric field polarization with the fiber's principal axis. To minimize the use of highly dispersive materials that would compromise the optimum pulse compression, the PCF output is collimated by an off-axis parabolic mirror and directed into a second compressor, consisting of a chirped-mirror pair (Ultrafast Innovations GmbH PC1611) with a nominal GDD of $-150~\text{fs}^2$ per bounce. The total number of bounces is adjusted to compensate for the normal dispersion introduced by the PCF, which varies with the fiber length. Following this final compression, nearly TL pulses are obtained. The input-to-output power efficiency is 78\%, regardless of the fiber length; thus, at maximum operating conditions, the system delivers an average power of 780~mW.

\begin{figure}[t]
\centering
\includegraphics[width=1\linewidth]{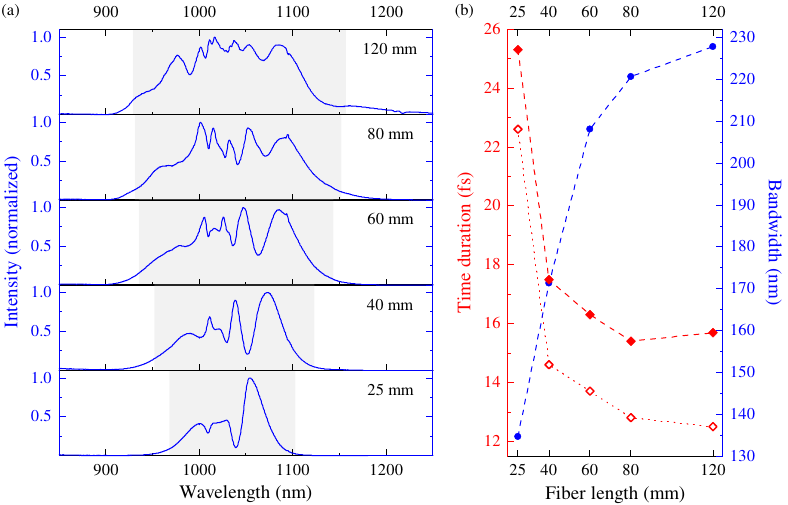}
\caption{Dependence on the spectral broadening for different photonic crystal fibers (PCF) lengths ranging from 25 to 120~mm. (a) Experimentally measured spectra. Gray areas represent the 4-rms spectral bandwidth. (b) Comparison between spectral and temporal properties: the plot displays the 4-rms spectral bandwidth (right axis, blue circles) together with the measured pulse duration (left axis, filled red diamonds) and the corresponding transform-limit duration (left axis, open red diamonds).}
\label{fig:length}
\end{figure}

First, the optimal fiber length for achieving the shortest pulse duration was investigated. Spectra were collected at the output of each PCF at maximum average power using an optical spectrum analyzer with a resolution of 2~nm. The results, reported in Figure \ref{fig:length}, show that increasing the fiber length causes the output spectrum to broaden from a 134.7~nm 4-rms bandwidth for the 25~mm fiber up to 227.8~nm for the 120~mm PCF. This increase is nearly linear for lengths between 25 and 60~mm, whereas saturation of the spectral broadening begins to appear beyond 60~mm. The experimental minimum pulse duration for each fiber length was characterized using the second harmonic generation (SHG) frequency-resolved optical gating (FROG) technique. The number of bounces on the chirped mirrors was adjusted to optimize GDD compensation and achieve the minimum pulse duration for each fiber, with fine-tuning performed using fused silica plates of varying thicknesses. Results in Figure \ref{fig:length}(b) indicate that a minimum pulse duration of 15.4~fs is achieved with the 80~mm fiber. This performance is attributed to its broader spectrum compared to the 25, 40, and 60~mm PCFs, which yield longer compressed pulses. The 120~mm-long fiber, despite showing a slightly larger spectrum, yielded a longer pulse duration compared to the 80~mm case. This is likely due to higher-order dispersion terms introduced by the PCF that cannot be fully compensated by the chirped-mirror pair. This behavior clearly indicates that spectral bandwidth alone is not a sufficient metric for pulse compression optimization, as higher-order phase distortions play a limiting role.

\begin{figure}[!ht]
\centering
\includegraphics[width=1\linewidth]{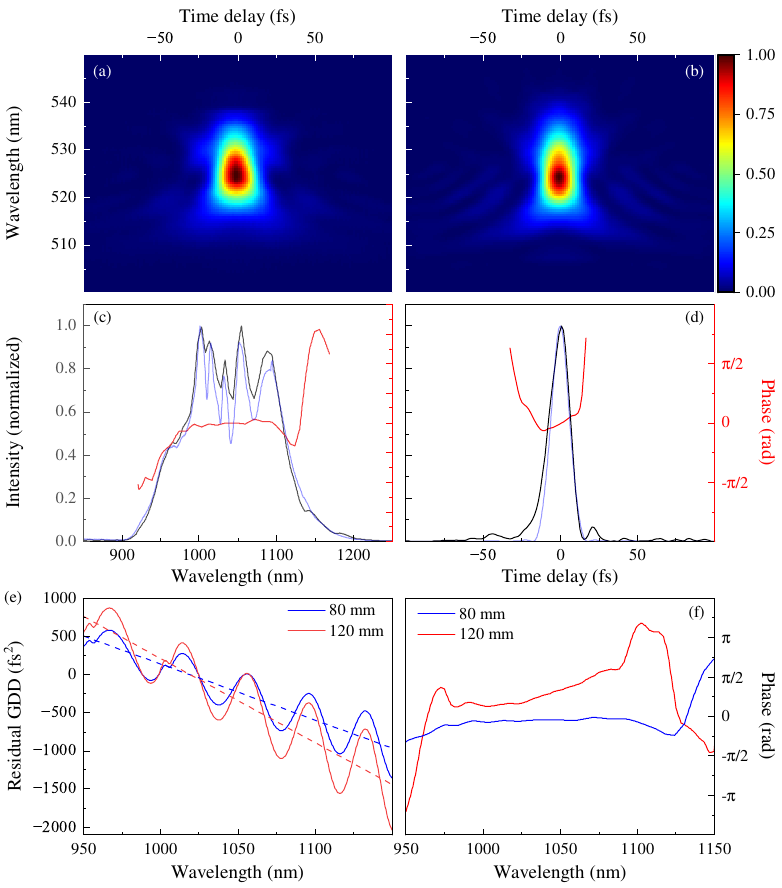}
\caption{Second harmonic generation frequency-resolved optical gating (FROG) characterization of the spectrally broadened pulse compressed to near-transform-limited (TL) 15.4~fs temporal duration. (a) Experimentally measured FROG trace. (b) Retrieved FROG trace. (c) Retrieved spectral intensity profile (black solid line), measured experimental spectrum (blue transparent line) and spectral phase (red solid line). (d) Retrieved temporal intensity profile (black solid line), TL temporal intensity profile (blue transparent line) and temporal phase (red solid line). The pulse exhibits a full-width at half-maximum duration of 15.4~fs close to the predicted TL duration of 12.8~fs. (e) Estimated residual group delay dispersion (GDD) affecting the pulse for the 80~mm (solid blue line) and 120~mm (solid red line) photonic crystal fibers (PCFs). Dashed lines represent the linear trend related to third-order dispersion (TOD). (f) Retrieved spectral phase  from the FROG algorithm of the 80~mm (blue solid line) and 120~mm (red solid line) PCFs.}
\label{fig:FROG}
\end{figure}

\begin{figure}[!ht]
\centering
\includegraphics[width=1\linewidth]{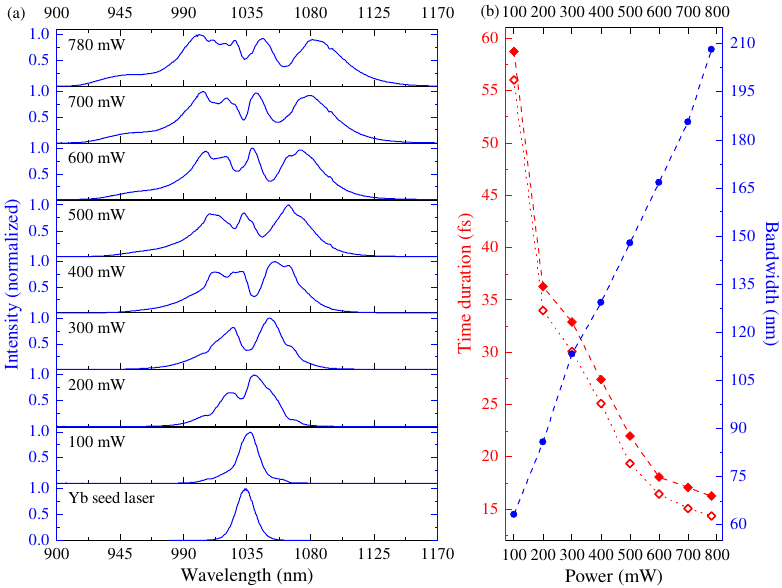}
\caption{Spectrum characterization as a function of the average power at the output of the 60~mm long photonic crystal fiber. (a) Evolution of the measured output spectra displayed as stacked plots for increasing power levels. (b) Quantitative analysis of the broadening and temporal duration: the 4-rms spectral bandwidth (right axis, blue circles) is reported along with the experimentally measured time duration (left axis, filled red diamonds) and the predicted transform-limit duration (left axis, open red diamonds).}
\label{fig:power}
\end{figure}

To further strengthen this argument, a quantitative analysis of the residual GDD and third-order dispersion (TOD) was performed for the 80~mm and 120~mm fibers. Figure \ref{fig:FROG} reports, in panels (a) and (b), the experimentally measured and algorithm-retrieved FROG maps relative to the 15.4~fs pulse obtained with the 80~mm PCF respectively; panels (c) and (d) display the retrieved spectral and temporal intensities (black solid lines) alongside their respective phases (red solid lines). The validity of the FROG retrieval is confirmed by the excellent agreement between the retrieved spectrum and the experimentally measured one (blue transparent line). The flatness of the spectral phase across the main bandwidth confirms effective GDD compensation. This is further supported by the temporal phase, which remains nearly constant over the central part of the pulse, indicating the successful removal of linear chirp. Furthermore, the resulting pulse is clean and essentially free of significant pre- or post-pulses, demonstrating high-quality compression. The pulse duration is very close to the TL value of 12.8~fs (blue transparent line).

Figures~\ref{fig:FROG}(e) and (f) compare the residual GDD affecting the pulse after compensation and the corresponding spectral phases for the 80~mm and 120~mm fibers. The residual GDD curves, obtained by summing the dispersion introduced by the PCF and the chirped mirrors, exhibit an overall linear trend as a function of wavelength, as highlighted by the dashed lines. This behavior is observed for both fiber lengths, but with a noticeably steeper slope in the case of the 120~mm-long fiber.

The presence of this linear trend indicates the contribution of uncompensated third-order dispersion (TOD). A linear fit of the GDD curve associated with the 120~mm PCF yields an estimated TOD of 6250~fs$^3$, which is significantly higher than the 4166~fs$^3$ retrieved for the 80~mm fiber (the difference in slope sign is due to the wavelength representation). This comparison confirms that the residual TOD increases with fiber length.

It is worth noting that the chirped mirrors provide a negligible contribution to the total TOD, indicating that the higher-order phase originates primarily from the PCF. The oscillations observed in the GDD curves are instead attributed to the chirped mirrors, with larger oscillation amplitudes in the 120~mm case due to the higher number of bounces required for GDD compensation.

The impact of this residual TOD is further confirmed by the spectral phase profiles shown in Figure~\ref{fig:FROG}(f). While the phase associated with the 80~mm fiber is nearly flat across the main spectral region, indicating effective compression, the phase of the 120~mm fiber exhibits significant deviations, which prevent full phase compensation and ultimately limit the achievable pulse duration.

\begin{figure}[t]
\centering
\includegraphics[width=1\linewidth]{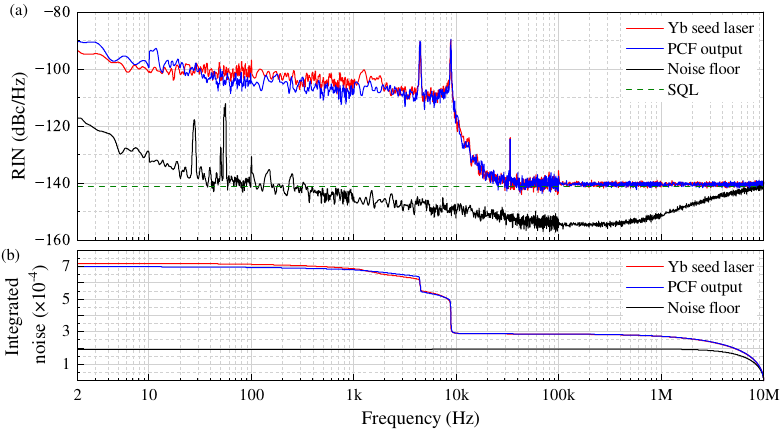}
\caption{Relative intensity noise (RIN) and integrated noise. (a) RIN spectra of the PCF output (blue), the Yb seed laser (red), the electronic noise floor (black), and the standard quantum limit (SQL) of -141~dBc/Hz, corresponding to a power of 50~\textmu W at 1034~nm (dashed green line). (b) Integrated rms intensity noise as calculated from RIN spectra.}
\label{fig:RIN}
\end{figure}

\section{60~mm fiber investigation}

The 60 mm fiber was systematically characterized to further investigate the interplay between spectral broadening and pulse compressibility as a function of the input power, and to assess the robustness of the identified optimal regime. Figure~\ref{fig:power} illustrates the spectral evolution at the output of the PCF for different power levels compared with the initial seed spectrum. From an initial 4-rms bandwidth of 29.6~nm, a 208~nm width is achieved at maximum power, following an essentially linear broadening trend. The compressed pulse duration exhibits the expected parabolic-like behavior, reaching a saturation regime at higher average powers. 
Notably, sub-20~fs pulses are successfully generated for average powers exceeding 600~mW. 

Finally, the short-term stability of the source was characterized via a relative intensity noise (RIN) measurement. The power spectral density of the photocurrent, generated by a photodiode (PDB450C, Thorlabs), was evaluated using an electrical spectrum analyzer. Figure \ref{fig:RIN} shows the RIN for the output of the PCF, along with that of the Yb seed laser and the electronic noise floor of the measurement setup. In both cases, the incident power on the photodiode was limited to 50~\textmu W, setting the standard quantum limit (SQL) to approximately -141~dBc/Hz. The RIN of the PCF closely follows that of the seed in the frequency range between 20~Hz and 10~MHz, with only a minor difference at low frequencies. The three peaks located at 4~kHz, 9~kHz, and 30~kHz are all attributed to the seed laser, where the 9~kHz peak corresponds to the relaxation oscillations. Above 40~kHz, both the seed and the PCF ultimately reach the SQL. The integrated RIN curves, shown in Figure \ref{fig:RIN}(b), confirm that the PCF noise is effectively limited by the one of the Yb laser, with an integrated noise level of $7\times10^{-4}$ in the range between 2~Hz and 10~MHz, which is slightly lower than the $7.2\times10^{-4}$ of the seed. 

These results further support the existence of an optimal operating regime, showing that while sufficient spectral broadening is required to reach the few-cycle regime, stable sub-20~fs operation can still be achieved over a range of conditions around the optimum.

\section{Conclusion}

We have identified an optimal nonlinear interaction regime for PCF-based compression of Yb femtosecond lasers, arising from the trade-off between spectral broadening and uncompensated higher-order phase, primarily third-order dispersion. Our results show that increasing the fiber length enhances the spectral bandwidth but also leads to the accumulation of higher-order phase distortions, which ultimately limit pulse compressibility.

Based on this analysis, we demonstrated sub-20-fs pulse generation from a commercial Yb laser using a compact PCF-based spectral broadening stage followed by chirped-mirror compression. A minimum near-transform-limited pulse duration of 15.4~fs was achieved with an 80~mm fiber, corresponding to the optimal interaction length.

Stable sub-20-fs operation was demonstrated above 600~mW average power, while relative intensity noise measurements confirmed that the system performance is mainly limited by the seed laser.

\begin{backmatter}
\bmsection{Acknowledgments} We thank the Laboratory Udyni of CNR-IFN and Politecnico di Milano for preliminary test of the compression system.
\bmsection{Funding} 
Funded by European Union, Next Generation EU, PNRR M4C2, investment 1.1, PRIN 2022, Project FANSECARS, ID 2022X29985, CUP D53D23002930006. E.A.A.P receives funding from the European Union (ERC, TREAT 101162914) and from the European Union’s Next-Generation EU Programme, Mission 4, Component 1, with the PRIN2022 project TRAPNE [CUP B53D23004290006].
\bmsection{Disclosures} The authors declare no conflicts of interest.

\end{backmatter}

\bibliography{biblio}

\end{document}